\documentclass[a4paper,11pt]{article}

\usepackage{jheppub}

\usepackage{epsfig,epsf}
\usepackage{graphicx}
\usepackage{amsfonts}
\usepackage{amsmath}
\usepackage{amsthm}
\usepackage{amssymb}

\def\be{\begin{equation}}
\def\lb#1{\label{#1}}
\def\ee{\end{equation}}
\def\bea{\begin{eqnarray}}
\def\eea{\end{eqnarray}}
\def\ba{\begin{array}}
\def\ea{\end{array}}
\def\dd{\partial}

\def\half{\frac{1}{2}}

\def\one#1{#1^{\raise5pt\hbox{$\scriptstyle\!\!\!\!1$}}\,{}}
\def\two#1{#1^{\raise5pt\hbox{$\scriptstyle\!\!\!\!2$}}\,{}}

\def\tilde{\widetilde}
\def\II{\hbox{{1}\kern-.25em\hbox{l}}}
\newcommand{\p}[1]{(\ref{#1})}
\newfont{\bbd}{msbm10 scaled\magstep1}

\def\B{\hbox{\bbd B}}

\newcommand{\cN}{\mathcal{N}}

\title{\boldmath A note on four-point correlators of  half-BPS operators 
in $\mathcal{N}=4$ SYM}

\author[a]{D. Chicherin}
\author[a,b,c]{and E. Sokatchev}

\affiliation[a]{LAPTH\,${}^{1}$\note{UMR 5108 du CNRS, associ\'ee \`a l'Universit\'e de Savoie}, 
Universit\'{e} de Savoie, CNRS,\\B.P. 110,  F-74941 Annecy-le-Vieux, France}
\affiliation[b]{Physics Department, Theory Unit, CERN,\\CH -1211, Geneva 23, Switzerland}
\affiliation[c]{Institut Universitaire de France,\\103, bd Saint-Michel
F-75005 Paris, France}

\emailAdd{chicherin@lapth.cnrs.fr}
\emailAdd{Emeri.Sokatchev@cern.ch}

\abstract{We calculate the four-point correlation function of half-BPS operators 
with weights 2, 3, 3, 4 in $\mathcal{N}=4$ SYM to two-loop order.
The OPE of this correlation function provides a nontrivial check of
the integrability conjecture for a class of three-point functions formulated in arXiv:1311.6404.
Our perturbative calculation exploits the supergraph formalism in $\mathcal{N} = 2$ harmonic superspace.}

\arxivnumber{1408.3527}

\notoc

\begin{document}
\maketitle
\flushbottom

\setcounter{footnote} 0


\setcounter{page}{1}

\section{Introduction}

In the paper~\cite{Vieira:2013wya}, which generalizes the results of \cite{Escobedo:2010xs,Gromov:2012vu} to the non-compact case, the 
three-point correlators of two half-BPS operators and 
one unprotected operator in the $SL(2)$ sector were studied in the one-loop approximation.
A conjecture was made, based on integrability, for the values of the corresponding structure constants. It was successfully confronted with the available perturbative results on the structure constants. 
In the absence of direct calculations of the relevant three-point correlators with unprotected operators,  use was made of the OPE of the two-loop four-point correlators of half-BPS operators, which produces sum rules for the structure constants. At the time when  the paper~\cite{Vieira:2013wya} was written, only results on four-point correlators of equal weights were available  \cite{Arutyunov:2003ad}. They allow one to test 
only sum rules that contain the squares of the structure constants. 
In order to perform a more detailed test of the integrability conjecture it is preferable to study more general sum rules
where products of two different structure constants appear. These can be obtained from two-loop four-point correlators of half-BPS operators of {\it different} weights. Here we present the calculation of  one such correlator whose OPE gives rise to non-symmetric sum rules. Our result has already been communicated to the authors of~\cite{Vieira:2013wya} who used it and found perfect agreement with their integrability prediction.

Our aim is to find  the two-loop approximation to the correlator 
\be \lb{corN4}
\mathcal{G}_{\mathcal{N} = 4}  = \langle \mathcal{O}^{(2)}(x_1,Y_1) \,\mathcal{O}^{(3)}(x_2,Y_2)\, \mathcal{O}^{(3)}(x_3,Y_3)\, \mathcal{O}^{(4)}(x_4,Y_4) \rangle
\ee
of  four half-BPS operators of weight $k$ defined by
$$
\mathcal{O}^{(k)} = Y_{I_1} \cdots Y_{I_k} \mathrm{Tr}\left( \Phi^{I_1} \cdots \Phi^{I_k} \right).
$$
Here $\Phi^{I}, I = 1, \cdots, 6$ are the six real scalars of $\mathcal{N} = 4$ SYM and $Y_{I}$ denotes an $SO(6)\sim SU(4)$ null vector, $(Y_i \cdot Y_i) = 0$ with $(Y_i \cdot Y_j) = (Y_j \cdot Y_i) = Y_i^{I} Y^{I}_{j}$. These auxiliary variables help us to keep track of the R symmetry indices. For our purposes it is more convenient to parametrize them by unconstrained complex $2\times 2$ matrices $y_{a'}^{\;a}$ (with $a = 1,2$, $a' = 1',2'$) transforming under the subgroup $SU(2)\times SU(2)' \subset SU(4)$. In these  terms  $(Y_i \cdot Y_j)= (y_i-y_j)^2 \equiv y_{ij}^2$ with $y^2 = - \half y_{a'}^{\;a} y^{a'}_{\;a}$.

$\mathcal{N} = 4$ superconformal symmetry imposes restrictions on the form of the quantum corrections to the four-point 
correlators of half-BPS operators~\cite{Eden:2000bk,Arutyunov:2003ae}.
For example, in the best known case of operators of weight $k=2$, which are
the lowest components of the $\mathcal{N} = 4$  stress-tensor multiplet, the correlator takes the general form
\be \lb{stm4}
 \langle \mathcal{O}^{(2)}(x_1,Y_1) \,\mathcal{O}^{(2)}(x_2,Y_2)\, \mathcal{O}^{(2)}(x_3,Y_3)\, \mathcal{O}^{(2)}(x_4,Y_4) \rangle = (\mbox{Born level}) + R_{\mathcal{N}=4}\, G(u,v)\,.
\ee
The loop corrections are encoded in the single function $G(u,v)=\sum_{n \geq 1} 
\lambda^{n} G_{n} (u,v)$ of the two conformal cross-ratios
$$
u = \frac{x_{12}^2 x_{34}^2}{x_{13}^2 x_{24}^2} \;\; , \;\; 
v = \frac{x_{14}^2 x_{23}^2}{x_{13}^2 x_{24}^2}\,.
$$
This function is expanded in powers of  the gauge coupling constant $g$ (or the `t Hooft coupling $\lambda = \frac{N g^2}{16 \pi^2}$). The quantum correction part of \p{stm4} is characterized by the universal prefactor
\begin{align}\label{R}
R_{\mathcal{N}=4} = u \frac{y_{12}^4 y_{34}^4}{x_{12}^4 x_{34}^4} + \frac{y_{13}^4 y_{24}^4}{x_{13}^4 x_{24}^4} 
+ v \frac{y_{14}^4 y_{23}^4}{x_{14}^4 x_{23}^4} + (v - u - 1) \frac{y_{12}^2 y_{13}^2 y_{24}^2 y_{34}^2}{x_{12}^2 x_{13}^2 x_{24}^2 x_{34}^2} + \notag\\
+ (1-u-v) \frac{y_{12}^2 y_{14}^2 y_{23}^2 y_{34}^2}{x_{12}^2 x_{14}^2 x_{23}^2 x_{34}^2} 
+ (u - v - 1) \frac{y_{13}^2 y_{14}^2 y_{23}^2 y_{24}^2}{x_{13}^2 x_{14}^2 x_{23}^2 x_{24}^2}
\end{align}
which supplies the necessary R symmetry and conformal weights at the four points. 
The non-trivial part of the correlator that cannot be fixed by symmetry alone is encoded in the functions $G_{n}$. In a generic correlator of four half-BPS operators of different weights $\langle \mathcal{O}^{(k_1)} \,\mathcal{O}^{(k_2)}\, \mathcal{O}^{(k_3)}\, \mathcal{O}^{(k_4)} \rangle$ (with $k_i \geq2$),  one needs an additional product of  propagators behind the universal prefactor \p{R} to supply the missing weights at each point. Due to the specific weight configuration in \p{corN4} there exists only one such propagator  structure with additional weights $(0,1,1,2)$. Consequently,  the correlator \p{corN4} involves a single function of the cross-ratios in full
analogy with \p{stm4}, 
\be \lb{GN4qc}
\mathcal{G}_{\mathcal{N}=4} =
(\mbox{Born level}) + R_{\mathcal{N}=4} \frac{y_{24}^2 y_{34}^2}{x_{24}^2 x_{34}^2} \sum_{n \geq 1} 
\lambda^{n} G_{n} (u,v)\,.
\ee

The result of this paper is the evaluation of the one- and two-loop contributions to \p{GN4qc}. Together with the Born approximation, they take the following form:
\begin{align}
(\mbox{Born level}) & = \frac{9\,\mathcal{C}_N}{4(2\pi)^{12}}\frac{y_{24}^2 y_{34}^2}{x_{24}^2 x_{34}^2} 
\left[ \frac{1}{2}\frac{y_{14}^4 y_{23}^4}{x_{14}^4 x_{23}^4} + \frac{1}{2}\frac{y_{12}^2 y_{13}^2 y_{24}^2 y_{34}^2}{x_{12}^2 x_{13}^2 x_{24}^2 x_{34}^2} + \frac{y_{12}^2 y_{14}^2 y_{23}^2 y_{34}^2}{x_{12}^2 x_{14}^2 x_{23}^2 x_{34}^2} 
+ \frac{y_{13}^2 y_{14}^2 y_{23}^2 y_{24}^2}{x_{13}^2 x_{14}^2 x_{23}^2 x_{24}^2} \right] \notag\\
G_{1} & = -\frac{9\, \mathcal{C}_N}{4(2\pi)^{12}} \,\Phi^{(1)}(u,v) \label{answer1}\\
G_{2} & = \frac{9\, \mathcal{C}_N}{4(2\pi)^{12}}
\biggl[ \frac{v}{2} \bigl[\Phi^{(1)}(u,v)\bigr]^2 + \Phi^{(2)}(u,v) + \frac{1}{u} \Phi^{(2)}(1/u,v/u) +
\frac{2}{v} \Phi^{(2)}(u/v,1/v)\biggr] \label{answer2}
\end{align}
with the color factor
\begin{align}\label{CN}
\mathcal{C}_N = (N^2-1)(N^2-4)(N^2-6)/N^2\,.
\end{align}
 The functions $\Phi^{(1)}$, $\Phi^{(2)}$ correspond to the so-called one- and two-loop ladder integrals. Their explicit expressions in terms of polylogarithms 
can be found in~\cite{Usyukina:1992jd,Usyukina:1993ch}. One can easily see that the crossing symmetry $2 \rightleftarrows 3$ is respected.

In order to check the correctness of our two-loop calculation we performed its OPE and compared it with 
known data on the anomalous dimensions.
The lowest dimension representation of $SU(4)$ in the overlap of the OPEs of the operators $\mathcal{O}^{(2)} \mathcal{O}^{(4)}$ 
and $\mathcal{O}^{(3)} \mathcal{O}^{(3)}$  is $[0,2,0]$. 
The twist $2$ operators in this channel are protected. 
Thus we have to consider the contribution of the twist $4$ scalar unprotected operators $\Theta_{\pm}$ whose anomalous dimensions
were calculated in \cite{Bianchi:2002rw}. Their values are in full agreement with the OPE of \p{corN4}. Another powerful check, as already mentioned,  is the confirmation of the integrability prediction in~\cite{Vieira:2013wya}. 

We find a similar result for the single-parameter family of correlators with $k \geq 1$
\begin{align} 
\mathcal{G}_{\mathcal{N} = 4}  = \langle \mathcal{O}^{(2)}(x_1,Y_1) \,\mathcal{O}^{(k+2)}(x_2,Y_2)\,
\mathcal{O}^{(k+2)}(x_3,Y_3)\, \mathcal{O}^{(2k+2)}(x_4,Y_4) \rangle\,, \notag
\end{align}
which is slight generalization of \p{corN4}. This is a subset of the two-parameter family of correlators
satisfying the next-next-to-extremality condition 
which has been considered in \cite{Uruchurtu:2011wh} 
at strong coupling\footnote{We are grateful to Hugh Osborn for drawing our attention to the paper~\cite{Uruchurtu:2011wh}. 
}. 
Like in \p{corN4}, 
its quantum corrections are encoded by a single function of the cross-ratios,
\begin{align} 
\mathcal{G}_{\mathcal{N}=4} =
(\mbox{Born level}) + R_{\mathcal{N}=4} \left(\frac{y_{24}^2 y_{34}^2}{x_{24}^2 x_{34}^2}\right)^k 
\sum_{n \geq 1} 
\lambda^{n} G_{n} (u,v)\,. \notag
\end{align}
Setting $k=1$ we get back to \p{corN4} and \p{GN4qc}. The calculation of this correlator essentially repeats that of \p{corN4}, so here we just quote the two-loop result in the leading color  limit $N \to \infty$. In the Born 
approximation we have  for $k \geq 2$
\begin{align}
(\mbox{Born level}) = \frac{4 N^{n+k}}{(8\pi^2)^{(n+k+2)}} (k+1)(k+2)^2
 \left(\frac{y_{24}^2 y_{34}^2}{x_{24}^2 x_{34}^2}\right)^k 
\left[ 2k \frac{y_{14}^4 y_{23}^4}{x_{14}^4 x_{23}^4} + 
\frac{y_{12}^2 y_{13}^2 y_{24}^2 y_{34}^2}{x_{12}^2 x_{13}^2 x_{24}^2 x_{34}^2} + \right. \notag\\ 
\left. +
(k+1) \frac{y_{12}^2 y_{14}^2 y_{23}^2 y_{34}^2}{x_{12}^2 x_{14}^2 x_{23}^2 x_{34}^2} 
+ (k+1) \frac{y_{13}^2 y_{14}^2 y_{23}^2 y_{24}^2}{x_{13}^2 x_{14}^2 x_{23}^2 x_{24}^2} + 
         k\frac{y_{13}^4 y_{24}^4}{x_{13}^4 x_{24}^4} 
       + k\frac{y_{12}^4 y_{34}^4}{x_{12}^4 x_{34}^4} \right]\,, \notag
\notag
\end{align}
while for $k = 1$ the last two terms should be omitted. The functional forms of the one- and two-loop corrections 
$G_1$ \p{answer1} and $G_2$ \p{answer2}
are unchanged, only the normalization factor is different,
\begin{align}
G_1, G_2:\ \   \frac{9\, \mathcal{C}_N}{4(2\pi)^{12}}\ \to\ \frac{8 (k+1) (k+2)^2 N^{2k+2}}{(8\pi^2)^{2k+4}} \,. \notag
\end{align}
The case $k = 0$ is special since the weights at all four points are equal and the correlator has full crossing symmetry.
The answer for $k = 0$ can be found for example in \cite{Arutyunov:2003ad}.

In the rest of this note we sketch some details of our perturbative calculation based on harmonic superspace Feynman rules.

\section{Calculation in $\mathcal{N} = 2$ harmonic superspace}

\subsection{Reduction $\mathcal{N}= 4 \to \mathcal{N} = 2$}

The $\mathcal{N} = 4$ SYM theory can be formulated  in  terms of  the $\mathcal{N}=2$ SYM multiplet and $\mathcal{N}=2$ hypermultiplet matter,  
\be \lb{action}
S_{\mathcal{N}=4 \;\mathrm{SYM}} = S_{\mathcal{N}=2 \;\mathrm{SYM}} + S_{\mathrm{HM}} \,.
\ee
The formulation  in $\mathcal{N}=2$ harmonic superspace (HSS)~\cite{Galperin:1984av,Galperin:2001uw} is completely off shell. This fact considerably facilitates the Feynman graph calculation using HSS supergraph techniques in combination with the 
Lagrangian insertion procedure~\cite{Howe:1999hz,Eden:2000mv} and the superconformal symmetry restrictions on the quantum corrections.
For a recent review of the method the reader is referred to Appendix A in~\cite{Eden:2010zz}.  

The evaluation of the $\mathcal{N} = 4$ correlator \p{corN4} is reduced  to the  calculation of the quantum corrections to the $\mathcal{N} = 2$  correlator
\be \lb{corN2}
\mathcal{G}_{\mathcal{N} = 2}  = \langle \tilde{\mathcal{Q}}^{(2)}(x_1,u_1) \,\mathcal{Q}^{(3)}(x_2,u_2)\, 
\mathcal{Q}^{(3)}(x_3,u_3)\, \tilde{\mathcal{Q}}^{(4)}(x_4,u_4) \rangle 
,\;
\mathcal{Q}^{(k)} = \mathrm{Tr}(q^{+})^k \;,\;\; \tilde{\mathcal{Q}}^{(k)} = \mathrm{Tr}(\tilde{q}^{+})^k
\ee
where the  half-BPS composite operators are constructed out of hypermultiplet  matter. 
The $\mathcal{N}=2$  hypermultiplet is described by a  Grassmann analytic superfield on HSS with coordinates 
$x^{\alpha\dot{\alpha}}_A,\theta^{+\alpha},\bar{\theta}^{+\dot{\alpha}},u^{\pm i}$, where the 
harmonic variables $u^{\pm i}$ form a matrix of $SU(2)$,
$$
|| u || \in SU(2) \;\;,\;\; u^{+i}u_{i}^{-} = 1 \;\;,\;\; \overline{u^{+i}} = u_{i}^{-} = \epsilon_{ij} u^{-j} 
$$
and $x_A^{\alpha\dot{\alpha}} = x^{\alpha\dot{\alpha}} - 4 i \theta^{\alpha (i} \bar{\theta}^{\dot{\alpha}j)} u^{+}_i u^{-}_j$. The harmonics transform under global $SU(2)$ (index $i=1,2$) and local $U(1)$ (weight $\pm 1$), thus they parametrize the harmonic coset $SU(2)/U(1) \sim S^2$. Grassmann analyticity means that only half of the Grassmann variables $\theta^{i \alpha}$, $\bar{\theta}^{i \dot{\alpha}}$ are involved, namely the harmonic projections  
$\theta^{+\alpha}=u^{+}_i \theta^{i\alpha}$, $\bar{\theta}^{+\dot{\alpha}}=u^{+}_i \bar{\theta}^{i\dot{\alpha}}$.
The on-shell hypermultiplet consists of an $SU(2)$ doublet of scalars $\phi^{i}(x)$ and singlet fermions $\psi_{\alpha},\bar{\kappa}^{\dot{\alpha}}$,
\begin{align}\label{HM}
q^{+}(x_A,\theta^{+},\bar{\theta}^{+},u) = \phi^{i}(x_A) u_{i}^{+} + \theta^{+\alpha}\psi_{\alpha}(x_A) 
+ \bar{\theta}^{+}_{\dot{\alpha}}\bar{\kappa}^{\dot\alpha}(x_A) 
+ 4 i \theta^{+} \sigma^{\mu} \bar{\theta}^{+} \dd_{\mu} \phi^{i}(x_A) u^{-}_{i}\,,
\end{align}
where the physical fields $\phi^i,\psi, \bar\kappa$  satisfy their free equations of motion. The HM can be lifted off shell by allowing it to depend on the harmonics in an arbitrary way, after which it becomes possible to write down an off-shell action~\cite{Galperin:1984av,Galperin:2001uw}. The HM  $q^{+}$ in \p{HM} is complex and 
$\widetilde{q}^{+}(x_A,\theta^{+},\bar{\theta}^{+},u)$ is its conjugate with the same analyticity.

The composite gauge-invariant operators in \p{corN2} are half-BPS in the sense that they depend on half of the Grassmann variables, $\mathcal{Q}^{(k)}(x_A, \theta^{+},\bar{\theta}^{+}, u)$ and  $\mathcal{\tilde Q}^{(k)}(x_A, \theta^{+},\bar{\theta}^{+}, u)$, just like their constituents $q^{+}$ and $\widetilde{q}^{+}$. In this paper we are interested in the four-point correlator  \p{corN2} of their lowest components at $\theta= \bar{\theta}= 0$. Nevertheless, as we explain below, the supersymmetric Feynman graph formalism that we use requires keeping track of the dependence on the chiral $\theta^+_\alpha$. 

The other ingredient in the $\cN=2$ formulation of $\cN=4$ SYM is the $\cN=2$ gauge multiplet. It is described by a chiral superfield (together with its antichiral conjugate) with the on-shell content 
\begin{align}\label{211}
W(x,\theta) = \varphi(x) + \theta^\alpha_i \lambda^i_\alpha(x) + \epsilon^{ij}\theta^\alpha_i \theta^\beta_j F_{\alpha\beta}(x)\,,
\end{align}
including a complex scalar $\varphi$, an $SU(2)$ doublet of chiral fermions $\lambda$ and the chiral (self-dual) half of the gauge field strength $F$.  

Like in the $\mathcal{N} = 4$ case, the quantum corrections to the correlator in \p{corN2} have the following partially non-renormalized form~\cite{Eden:2000bk}
\be\label{29}
\mathcal{G}_{\mathcal{N} = 2}  = (\mbox{Born level}) + 
\frac{(24)(34)}{x_{24}^2 x_{34}^2} R_{\mathcal{N}=2} \;G(u,v)
\ee
with the universal prefactor
\be \lb{R2}
R_{\mathcal{N}=2} = u \frac{(12)^2(34)^2}{x_{12}^4 x_{34}^4} + \frac{(13)^2(24)^2}{x_{13}^4 x_{24}^4} 
+ (v-u-1) \frac{(12)(13)(24)(34)}{x_{12}^2 x_{13}^2 x_{24}^2 x_{34}^2}\,.
\ee
We denote the contractions of a pair of $SU(2)$ harmonics {(both having weight $+1$)} referring to points $1$ and $2$ as follows
$$
(12) = - (21) = u^{+i}_{1} \epsilon_{ij} u^{+j}_2\,. 
$$
Note that each term in \p{R2} has uniform harmonic and conformal weights $+2$ at each point. The prefactor in the quantum correction part in \p{29} supplies the additional weights needed for the correlator  \p{corN2}.

It is important to realize that the loop correction function $G(u,v)$ in the $\mathcal{N} = 2$ correlator \p{29} is exactly the same as in the $\mathcal{N} = 4$ one \p{GN4qc}. This can be shown by 
a reduction $\mathcal{N}= 4 \to \mathcal{N} = 2$,  as explained in \cite{Arutyunov:2002fh,Arutyunov:2003ae} in terms of 
harmonics and in \cite{Eden:2011ku} using $y$-variables. Here we resort to the latter formalism.
The field 
strength multiplet $\mathcal{W}_{\mathcal{N} = 4}$ is reduced to its  $\mathcal{N} = 2$ projections, the chiral field 
strength multiplet $W(x,\theta)$ and the analytic HMs $q^+, \tilde q^+$, by means  of the $SU(4)$ raising operators 
$D_{a}^{a'}$: $D_{a}^{a'} y_{b'}^{\;b} = \delta^{b}_{a} \delta^{a'}_{b'}$. After acting with it on $\mathcal{W}_{\mathcal{N} = 4}$, we set $y^{\;1}_{1'},y^{\;1}_{2'},y^{\;2}_{2'} \to 0$, $y^{\;2}_{1'} \to \mathbf{y}$, thus reducing the $SU(4)$ harmonics to $SU(2)$ harmonics $(1,\mathbf{y}) = u^{+}_i$. The HMs are identified with the following projections: $\mathcal{W}_{\mathcal{N} = 4} \ \to \ q^+$, $D_{1}^{2'}\mathcal{W}_{\mathcal{N} = 4} \ \to \ \tilde q^+$ while the $\cN=2$ field strength is given by $D_{1}^{1'}\mathcal{W}_{\mathcal{N} = 4} \ \to \  W$. In this way we obtain \p{29} as a particular projection of \p{GN4qc}, 
$$
\mathcal{G}_{\mathcal{N} = 2} = \frac{1}{48}\bigl( D_{1}^{2'} |_1 \bigr)^2 \bigl( D_{1}^{2'} |_4 \bigr)^4 \; \mathcal{G}_{\mathcal{N} = 4} \,.
$$ 
One can easily check that this differential operator reduces the prefactor $ R_{\mathcal{N}=4}$ to $ R_{\mathcal{N}=2}$.

We chose the particular projection \p{corN2} since it is related in a very simple way to the full $\mathcal{N} = 4$ correlator \p{corN4}
we are interested in. Indeed, in order to reconstruct $\mathcal{G}_{\mathcal{N} = 4}$ in \p{corN4} from   $\mathcal{G}_{\mathcal{N} = 2}$ in \p{corN2}, 
we simply replace $R_{\mathcal{N} = 2}$ by $R_{\mathcal{N} = 4}$ and substitute the contractions of $SU(2)$ harmonics by $SO(6)$ harmonics, keeping the same loop correction function $G(u,v)$. 
Moreover, as we show below, due to the particular choice of $\mathcal{N} = 2$ half-BPS operators in   \p{corN2}
the number of relevant topologies of the contributing Feynman diagrams is rather small,
so this projection can be calculated quite easily.

\subsection{Lagrangian insertion and Feynman rules}

We calculate the quantum corrections to the correlator by means of the Lagrangian insertion procedure~\cite{Howe:1999hz,Eden:2000mv}. Here we give a brief outline. Consider the four-point correlator of some operators $\mathcal{O}(x,u)$ (not necessarily the same)
\be \lb{corr}
\mathcal{G} = \langle \mathcal{O}(1) \mathcal{O}(2) \mathcal{O}(3) \mathcal{O}(4) \rangle =
(\mbox{Born level})  + g^2\mathcal{G}_{\rm 1-loop} + g^4\mathcal{G}_{\rm 2-loop} + \cdots \,.
\ee
 The first quantum correction
$\mathcal{G}_{\rm 1-loop}$ is given by the derivative ${\dd\,\mathcal{G}}/{\dd \,g^2}|_{g=0}$. After rescaling the gauge and matter superfields in the $\mathcal{N} = 4$ action by the gauge coupling constant $g$, the latter appears only in front of the chiral gauge filed strength  $W(x,\theta^{i\alpha})$ in the $\mathcal{N} = 2$ SYM Lagrangian 
\begin{align}\label{}
\mathcal{L}_{\mathcal{N}=2 \;\mathrm{SYM}} = \frac{1}{4g^2} \mathrm{Tr} \,W^2\,.
\end{align}
The differentiation with respect to $g^2$ in the path integral brings down an insertion of the $\mathcal{N} = 2$ SYM action
\begin{align}\label{216}
S_{\mathcal{N}=2 \;\mathrm{SYM}} = 
\int d^4 x  d^4\theta \,\mathcal{L}_{\mathcal{N}=2 \;\mathrm{SYM}}(x,\theta) \,,
\end{align}
and we obtain
\be \lb{insert}
\frac{\dd\,\mathcal{G}}{\dd \,g^2} = 
- \frac{i}{g^2} \int d^4 x_5 d^4 \theta_5
\langle \mathcal{O}(1) \mathcal{O}(2) \mathcal{O}(3) \mathcal{O}(4)\ 
\mathcal{L}_{\mathcal{N}=2 \;\mathrm{SYM}}(x_5,\theta_5)\rangle\,.
\ee
This means that the one-loop correction in \p{corr} is given by the five-point correlator in \p{insert} calculated at Born level and integrated over the insertion point $x_5, \theta_5$. This correlator is itself $\sim g^2$, therefore we can safely set $g=0$ in \p{insert} to obtain $\mathcal{G}_{\rm 1-loop}$.
In the same way the double differentiation of \p{corr} results in a double Lagrangian insertion, still
calculated at Born level and integrated over the two insertion points. This gives  the second quantum correction $\mathcal{G}_{\rm 2-loop}$.

We remark that the $\mathcal{N} = 2$ SYM filed strength $W$ carries R charge\footnote{The $\cN=2$ R charge is the $U(1)$ factor in $U(1)\times SU(2) \subset SU(4)$, obtained by reduction from $\cN=4$.} $+2$ in units in which the R charge of the chiral Grassmann variable $\theta_\alpha$ equals $+1$, so that the action \p{216} is chargeless.  On the other hand, the HM has no R charge, therefore the five-point correlator with the Lagrangian insertion in \p{insert} has R charge $+4$. We conclude  that the five-point correlator  in \p{insert} must be nilpotent, $\sim \theta^4_5$. So, although we are interested in the bosonic four-point correlator \p{corr} (i.e. the lowest component of a super-correlator), the insertion formula 
requires a nontrivial dependence on the chiral Grassmann variables. At the same time, we can set all antichiral $\bar\theta=0$.

As we explain below, in our perturbative calculation up to order $g^4$ the gauge part of the action \p{action} manifest itself solely as insertion points 
in the Feynman diagrams. The gauge self-interaction is not relevant at this order.
The HM action $S_{\mathrm{HM}}$ in \p{action} gives rise to the matter propagator and the gauge-matter interaction vertex.

The hypermultiplet propagator in the adjoint representation of the gauge group $SU(N)$,\footnote{The generators of the fundamental irrep of the color group $SU(N)$ are normalized as ${\rm Tr} (t_a t_b) = \half \delta_{ab} $.} evaluated at $\bar\theta=0$, takes the following very simple form:
\be \lb{hmp}
\begin{array}{c}\includegraphics[width = 3 cm]{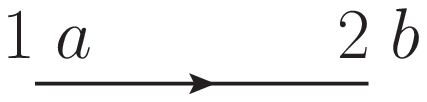}\end{array} 
= \langle \tilde{q}^{+}_a( x_1 , \theta^{+}_1,0,u_1) \, q^{+}_b ( x_2 , \theta^{+}_2,0,u_2) \rangle = \frac{(12)}{(2\pi)^2 x_{12}^2} \delta_{ab}\,.
\ee
It consists of an ordinary scalar propagator $1/x^2_{12}$ and a harmonic factor $(12)$ which keeps track of the isodoublet indices of the scalar fields. 
Another basic building block in the Feynman graphs is the Born level three-point function 
of two analytic hypermultiplet superfields and one chiral superfield strength, the so-called $\mathrm{T}$-block.
It is a rational function of the superspace coordinates calculated in~\cite{Eden:2000mv}.
For our purposes we will need it only at $\bar{\theta}^{+}_{1,2} = 0$ and $u_1^{\pm} = u_2^{\pm}$, which simplifies it significantly, 
\be \lb{Tblock}
\mathrm{T}^{abc}_{152} = \langle \tilde{q}^{+}_a(1) W_b(5) q^{+}_c(2) \rangle_{u_1=u_2} = \begin{array}{c}\includegraphics[width = 3.5 cm]{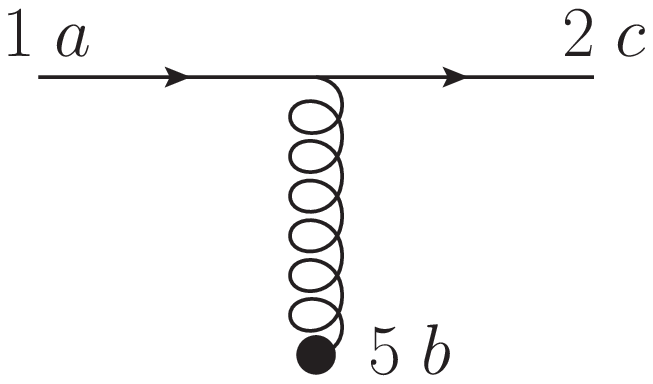} \end{array} 
= - \frac{2g f_{abc}}{(2\pi)^4}\frac{(\rho_1-\rho_2)^2}{x_{12}^2} 
\ee
where
$\rho^{\dot\alpha}_r = \bigl(\theta^i_{5} (u_r)^{+}_i - \theta^{+}_r \bigr)_{\alpha} \left(x_{5r}^{-1}\right)^{\alpha \dot\alpha}$. As remarked above, this object carries the R charge $+2$ of the chiral field strength $W$, therefore it is nilpotent, $\sim \theta^2$.

\subsection{One loop}

In order to obtain the one-loop correction to the four-point correlator we apply the insertion formula once, so we need to calculate the Born level five-point correlator  
\be \lb{Gg2}
\mathcal{G}^{\mathrm{ins}}_{g^2} = \langle \tilde{\mathcal{Q}}^{(2)}_1 \mathcal{Q}^{(3)}_2 \mathcal{Q}^{(3)}_3 \tilde{\mathcal{Q}}^{(4)}_4 
\frac{1}{g^2}\mathrm{Tr} W_5^2 \rangle_{\mathrm{Born}} = (42)(43) \,\Theta_{5}\, G_{g^2}(x_1,x_2,x_3,x_4,x_5)
\ee
at $\bar{\theta}_5 = 0$.
The presence of the nilpotent invariant
\be \lb{theta5}
\Theta_{5}  = \theta_5^4 \frac{R_{\mathcal{N}=2}}{\prod_{i=1}^4 x_{i5}^2} x_{12}^2 x_{34}^2 x_{13}^4 x_{24}^4 + \mbox{($\theta^+$ terms)} 
\ee
is a corollary of $\cN=2$ superconformal symmetry  \cite{Eden:2000mv}. It carries harmonic weights $+2$ at each point and R charge $+4$. 
The harmonic prefactor $(42)(43)$ completes the harmonic weights to $+2,+3,+3,+4$ at points $1,2,3,4$, respectively. 

We point out an important property of the correlator \p{Gg2}: it involves only harmonics $u^+_i$ with positive harmonic weight. This follows from one of the defining properties of the half-BPS operators $\mathcal{Q}^{(k)}(x,u)$, namely, they must be polynomials in $u^+_i$ of degree $k$. This property is called H-analyticity and reflects the fact that the operators are described by finite-dimensional representations of the R symmetry group.

\begin{figure}[tbp]
\begin{center}
\begin{tabular}{ccc}
\includegraphics[width =3 cm]{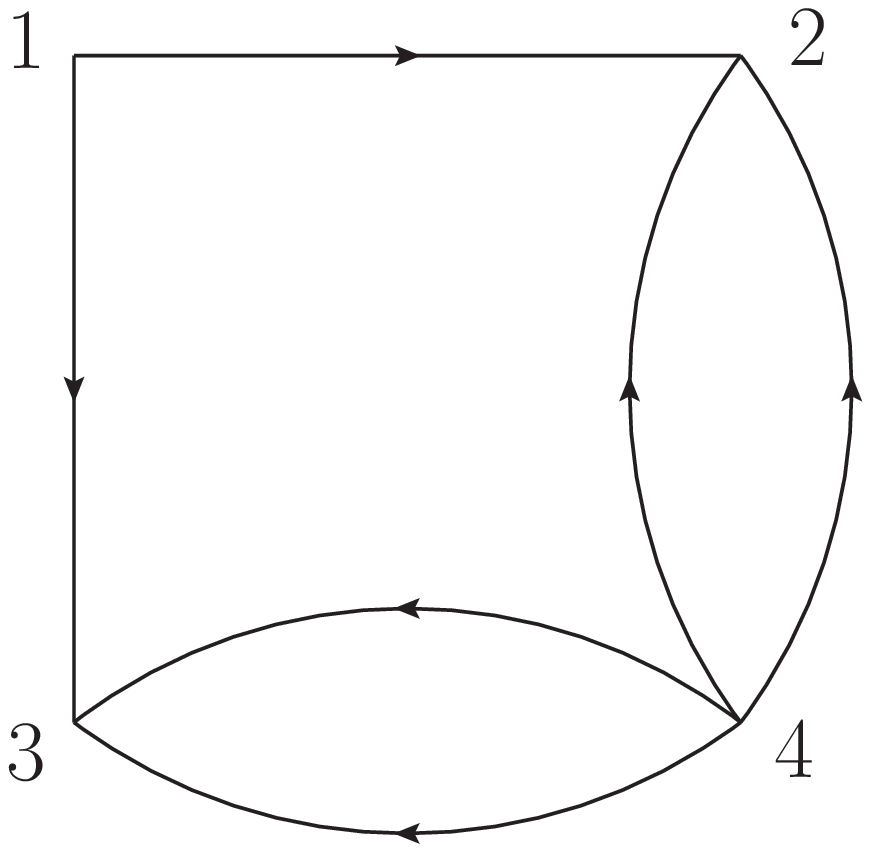} \; & \; 
\includegraphics[width =3 cm]{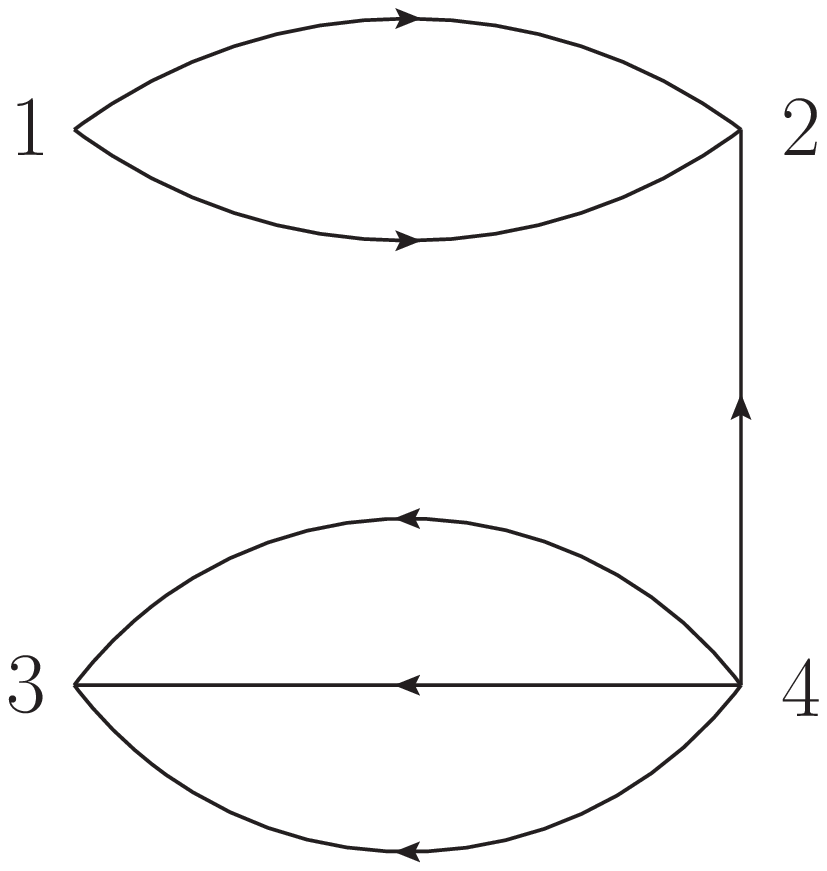} \; &\;
\includegraphics[width =3.5 cm]{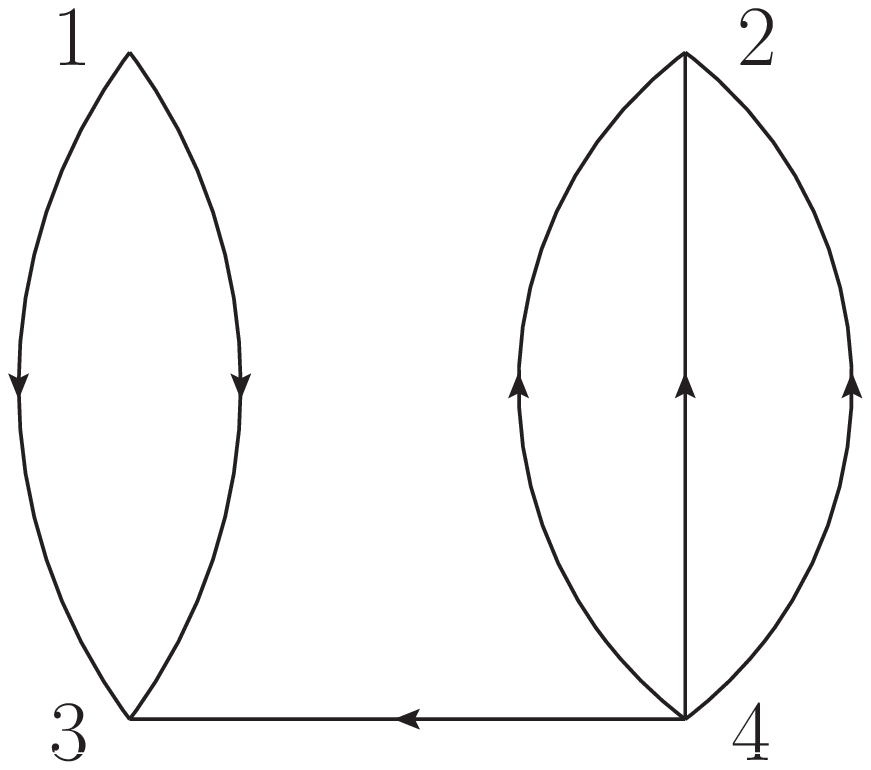}  \\
(F1) & (F2) & (F3)
\end{tabular}
\end{center}
\caption{\label{Fig1}  Three types of frames made of HM propagators.}
\end{figure}

The relevant Feynman diagrams can be obtained from three types of free matter line frames depicted in figure \ref{Fig1}
by inserting  an additional chiral vertex $1/g^2 \,\mathrm{Tr} W^2(5)$ and connecting it with the frame  lines by  gauge/matter interactions \p{Tblock}. An example is shown in eq.~\p{fig} below. 

We wish to simplify the Feynman graph calculation as much as possible. The general form of the 
correlator \p{Gg2} with a single Lagrangian insertion suggests to partially identify the harmonics, 
\be \lb{hequal}
u^{\pm}_1 = u^{\pm}_2 \;\;,\;\; u^{\pm}_3 = u^{\pm}_4\,.
\ee  
As a result, the nilpotent invariant \p{theta5} simplifies significantly,
$$
\Theta_{5}|_{u^{\pm}_1 = u^{\pm}_2 ,  u^{\pm}_3 = u^{\pm}_4} = (13)^2 (\rho_1 - \rho_2)^2 (\rho_3 - \rho_4)^2\,.
$$
However, such an identification  would also make the harmonic prefactor in \p{Gg2} vanish. Indeed, the diagram  in \p{fig} has an extra free matter line $\sim (34) \to 0$ (see \p{hmp}). 
We have to be more cautious. Following~\cite{Arutyunov:2003ad}, we first pull the harmonic prefactor $(42)(43)$ out of the diagrams and 
only afterwards  perform the identification \p{hequal}. In \p{Gg2} this corresponds to factoring out the free matter lines 2-4 and 3-4 prior to the identification. Note that  we cannot impose further restrictions on the harmonics, otherwise either $\Theta_{5}$ or the harmonic prefactor in \p{Gg2} will vanish. 

After the separation of the harmonic prefactor $(42)(43)$ coming from the free hypermultiplet lines \p{hmp},
we have to insert the gauge/matter interaction vertices into the remaining lines connecting the pairs of outer points 1 and 2, and 3 and 4. This prevents the diagram with the topology  (F1) from vanishing upon the identification \p{hequal}. The second topology  (F2)  involves an extra free line 3-4, so it does not contribute. 
The last topology (F3), after the insertion of the gauge/matter vertices, vanishes for color reasons.  

So, the unique contribution,  coming from the diagram with topology (F1), takes the form
\begin{align}\label{fig}
\begin{array}{c}
\includegraphics[width =3 cm]{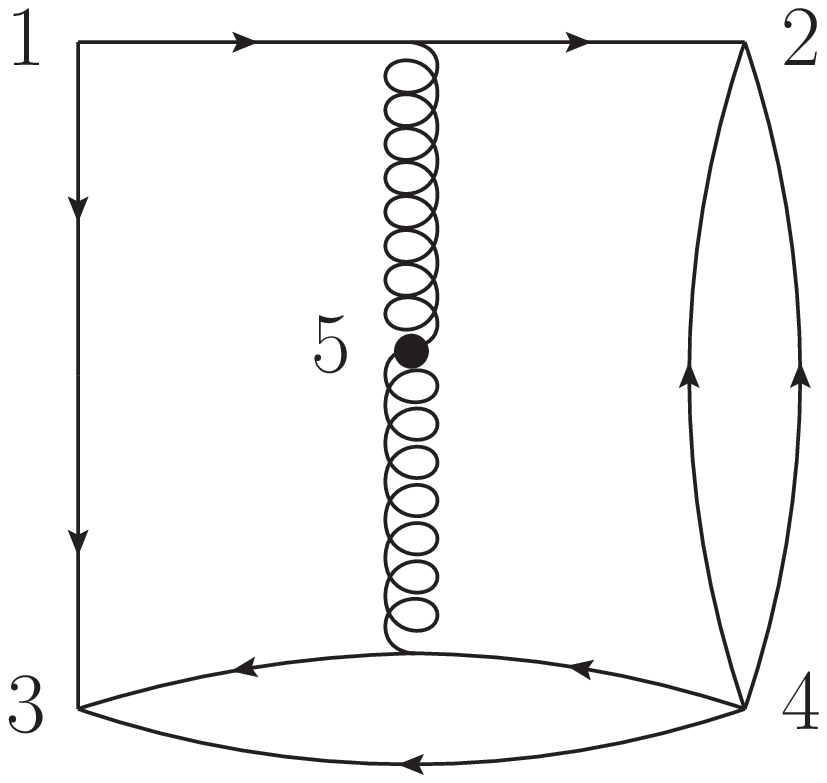}
\end{array} \sim (42)(43) \times (13)(42) \mathrm{T}_{453} \mathrm{T}_{152} \,.
\end{align}
It contains the required harmonic prefactor from \p{Gg2}. The identification \p{hequal} in the rest of the diagram immediately reproduces 
the nilpotent invariant $\Theta_5$, 
$(13)(42) \mathrm{T}_{453} \mathrm{T}_{152} \to - (13)^2 (\rho_1 -\rho_2)^2 (\rho_3 -\rho_4)^2 = -\Theta_5$.
Thus the calculation is almost trivial. We just need to take into account the color and symmetric factors to obtain
$$
G_{g^2} = \frac{9g^2\mathcal{C}_N N}{(2\pi)^{16}}\frac{1}{x_{12}^2 x_{13}^2 x_{24}^4 x_{34}^4} 
$$
with $\mathcal{C}_N$ defined in \p{CN}.

Finally, we apply the insertion formula
\be \lb{5ins}
\mathcal{G}_{g^2}
= - \frac{i}{4} \int \mathrm{d}^4 x_5 \mathrm{d}^4 \theta_5  \mathcal{G}^{\mathrm{ins}}_{g^2}\,.
\ee
The Grassmann integration in \p{5ins} picks out the first term  of the nilpotent invariant  \p{theta5}, 
and the $x$-space integration produces the standard one-loop ladder integral $\Phi^{(1)}(u,v)$, as stated in \p{answer1}. 

In the calculation above we retained only the Feynman diagrams that contribute to the correlator
after the identification of harmonics. However, the calculation 
can also be performed without this identification. In that case one has to deal with a number
of auxiliary Feynman diagrams which involve harmonics $u^{-}$ as well. According to the general form \p{Gg2} of the five-point nilpotent correlator and the underlying property of H-analyticity, the harmonics $u^{-}$ must drop out after summing up all diagrams. As a check, one can carry out the complete calculation and track the cancellation of the harmonics $u^{-}$ in the sum.

\subsection{Two loops}
According to the insertion procedure, the two-loop correction to the four-point correlator is expressed through the Born-level six-point correlator 
with two $\mathcal{N} = 2$ SYM Lagrangian insertions,
\be \lb{Gg4}
\mathcal{G}^{\mathrm{ins}}_{g^4} = \langle \tilde{\mathcal{Q}}^{(2)}_1 \mathcal{Q}^{(3)}_2 \mathcal{Q}^{(3)}_3 \tilde{\mathcal{Q}}^{(4)}_4 
\frac{1}{g^2}\mathrm{Tr} W_5^2 \,\frac{1}{g^2}\mathrm{Tr} W_6^2 \rangle_{\mathrm{Born}} = (42)(43) \,\Theta_{5,6}\, G_{g^4}(x_1,x_2,x_3,x_4,x_5,x_6)
\ee
at $\bar{\theta}_5 = \bar{\theta}_6 = 0$.
The new nilpotent invariant $\Theta_{5,6}$ is again a corollary of $\cN=2$ conformal supersymmetry   \cite{Eden:2000mv}. It has the form
\begin{align}\label{Theta56}
\Theta_{5,6} = \Big[ \theta_5^4 \theta^4_6 \, x_{12}^2 x_{34}^2 x_{13}^4 x_{24}^4 R_{\mathcal{N}=2} 
 +\ldots + \left(\theta^{+}_1\right)^2 \left(\theta^{+}_2\right)^2 
\left(\theta^{+}_3\right)^2 \left(\theta^{+}_4\right)^2 x_{56}^4 \Big] \frac{x_{56}^4}{\prod_{i=1}^4 x_{i5}^2 x_{i6}^2} 
\end{align}
and carries harmonic weights $+2$ at each point and R charge  $+8$. 
The harmonic prefactor $(42)(43)$ completes the harmonic weights to $+2,+3,+3,+4$ at points $1,2,3,4$, respectively. 
The dynamical information is contained in the  conformally covariant function of the six $x$-space coordinates  $G_{g^4}$.

\begin{figure}[tbp]
\hspace{-1.0 cm}
\begin{tabular}{ccccc}
\includegraphics[width =3.5 cm]{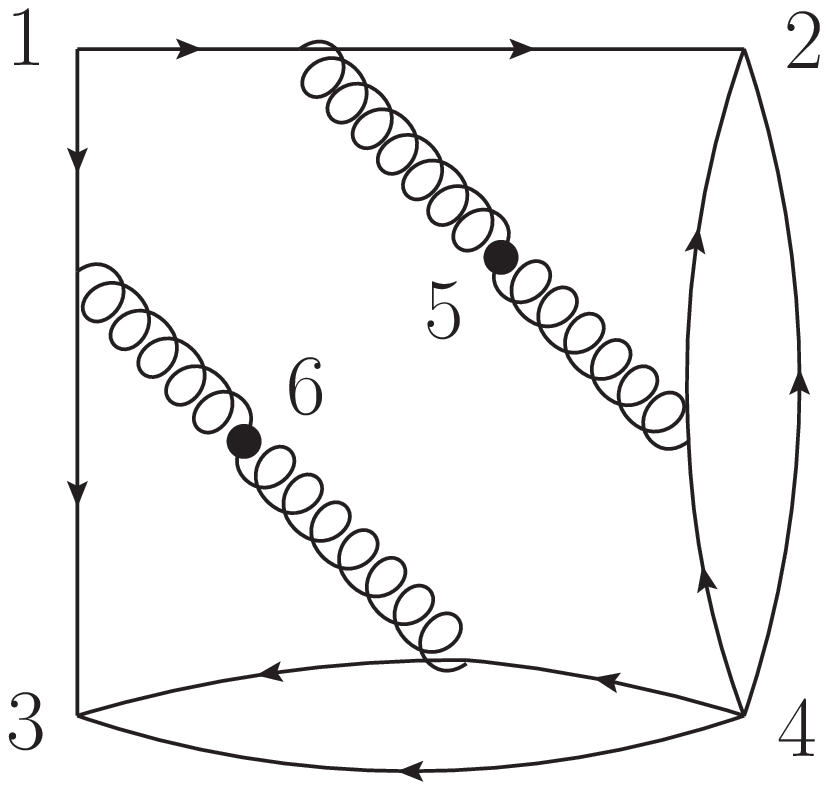} &
\includegraphics[width =3 cm]{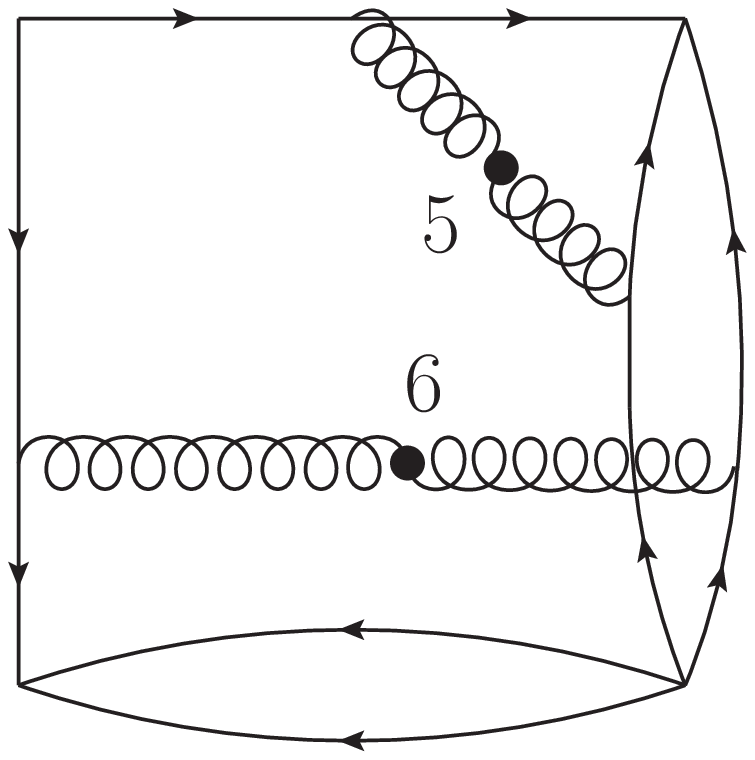} &
\includegraphics[width =3 cm]{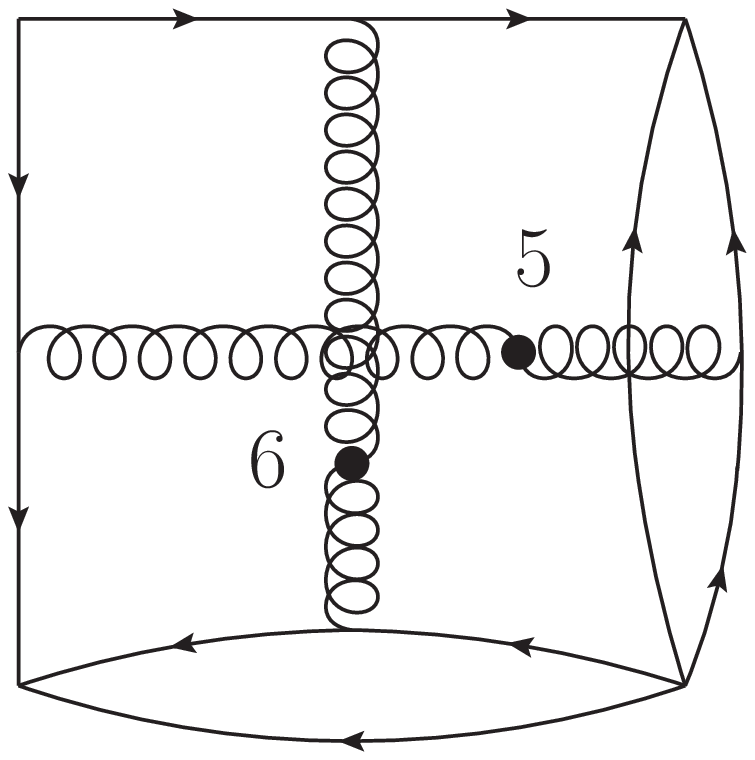} &
\includegraphics[width =3 cm]{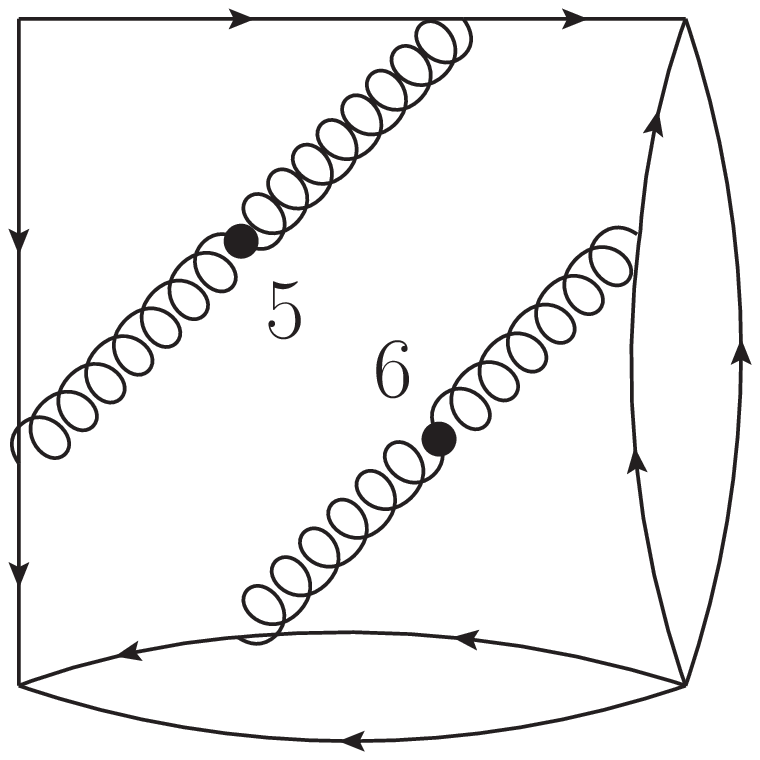} &
\includegraphics[width =3.5 cm]{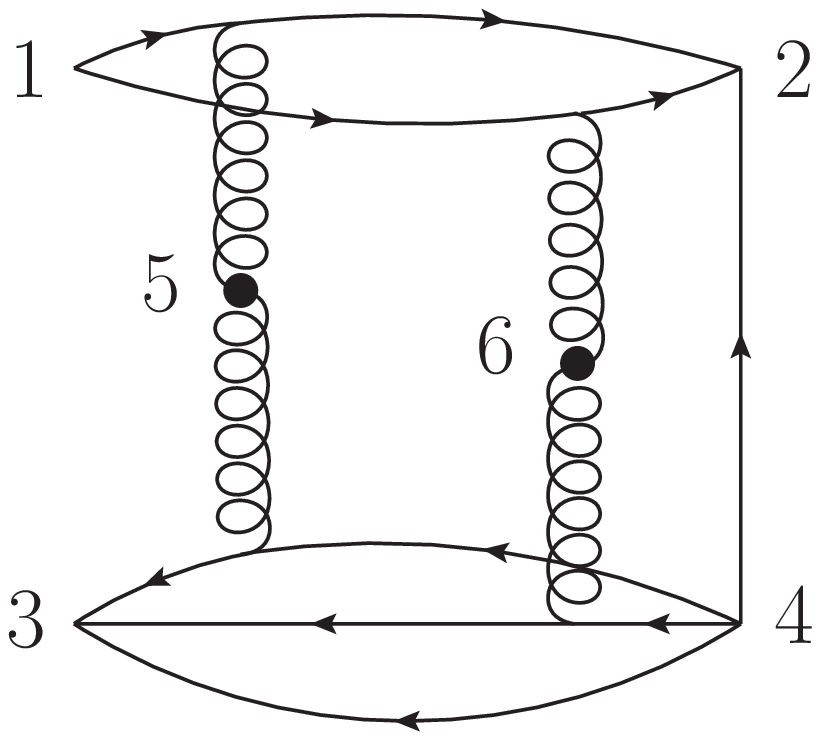} \\
(A) & (B) & (C) & (D) & (E)
\end{tabular}
\caption{\label{Fig2} $\mathcal{N} = 2$ supergraphs contributing at two-loop order.}
\end{figure}

We are interested in the first component of $\Theta_{5,6}\sim \theta^4_5 \theta^4_6 + \ldots $ needed for the integration over the insertion points. However, in order to simplify the Feynman graph calculation it proves convenient to first choose the frame $\theta_5 = \theta_6=0$
in which the invariant \p{Theta56} is reduced to its last component. The advantage of this frame is that the analytic $\theta^+$ in the last term carry harmonic weight $+2$ at each point. The remaining weights are supplied by the explicit harmonics prefactor $(42)(43)$ in \p{Gg4}. Then it becomes possible to identify three pairs of harmonics, 
$u^{\pm}_1 = u^{\pm}_2 = u^{\pm}_3$. We can do even better, as we did in the one-loop case, by first pulling the residual prefactor $(42)^2$ out of each Feynman diagram. After that we are allowed to set all the harmonics equal,
\be \lb{allhequal}
u^{\pm}_1 = u^{\pm}_2 = u^{\pm}_3 = u^{\pm}_4\,.
\ee
Indeed, neither the harmonic-independent function $G_{g^2}$ nor the invariant $\Theta_{5,6}$ vanish after this identification.  
This enables us to considerably reduce the number of diagrams we need to deal with. Indeed, each free HM line 
(i.e. a propagator without an insertion of the gauge interaction vertex) connecting a pair of external points $i$ and $j$ is proportional to $(ij)$, see \p{hmp}.
Thus all diagrams with three or more free HM lines left vanish after the identification of all harmonics \p{allhequal}, so we discard them and consider only 
diagrams with no more than two free lines, which give rise to the harmonic prefactor $(42)^2$. 
Then we note that the diagrams with subgrahps like
$$
\begin{array}{c}\includegraphics[width =3.5 cm]{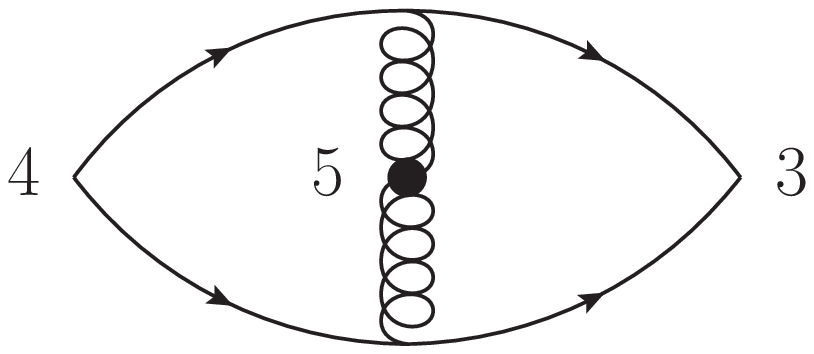}\end{array} = \mathrm{T}_{453} \mathrm{T}_{453} \sim (\rho_3 -\rho_4)^4 = 0
$$
vanish due to the odd nature of $\rho^{\dot\alpha}$. 
We arrive at the set of relevant Feynman diagrams with non-vanishing  color factors depicted in figure \ref{Fig2}.

They are constructed out of $\mathrm{T}$-blocks \p{Tblock} inserted in the mater line frames (F1), (F2), (F3) 
and their nontrivial contributions are equal to
\begin{align*}
&(A) = \rho_2^2\sigma_3^2 \tau_{14} \;\; , \;\; (B) = \rho_1^2 \sigma_3^2 \tau_{24} \;\;,\;\; 
(C) = \rho_1^2 \sigma_2^2 \sigma_3^2 \rho_4^2 + \sigma_1^2 \rho_2^2 \rho_3^2 \sigma_4^2 
+ 16 (\sigma_1 \sigma_2)(\rho_1 \rho_3)  (\rho_2 \rho_4) (\sigma_3 \sigma_4) \;\;,\\
&(D) = \rho_1^2 \sigma_4^2 \tau_{23} \;\;,\;\; (E) = \tau_{12} \tau_{34}
\end{align*}
where  $\sigma$ denotes the analog of $\rho$ for the second insertion point $6$, 
$\sigma^{\dot\alpha}_r = \bigl(\theta^{i}_{6} (u_r)^{+}_{i} - \theta^{+}_r \bigr)_{\alpha} \left(x_{6r}^{-1}\right)^{\alpha \dot\alpha} $, and the shorthand notation 
$
\tau_{rs} = \rho_r^2 \sigma_s^2 + \rho_s^2 \sigma_r^2 + 4 (\rho_r \rho_s) (\sigma_r \sigma_s)$ is used. 
Here we do not display the overall factor ${(2g)^4}/{(2\pi)^{20}}$
coming from the $\mathrm{T}$-blocks \p{Tblock} and the HM propagators, as well as the  propagator factors 
$\frac{1}{x_{12}^2 x_{13}^2 x_{24}^4 x_{34}^4}$ in diagrams (A), (B), (C), (D) and $\frac{1}{x_{12}^4 x_{24}^2 x_{34}^6}$ 
in diagram (E).
Let us also recall that we omit the harmonic prefactor $(42)^2$ which we singled out before the identification \p{allhequal}.

The symmetry and color factors for the above diagrams are as follows:
$$
\mathcal{C}_A = \mathcal{C}_D = \frac{9}{2}\mathcal{C}_N N^2 \;\;,\;\;
- \mathcal{C}_B = \mathcal{C}_C = \mathcal{C}_E =  \frac{9}{4}\mathcal{C}_N N^2  
$$
with $\mathcal{C}_N$ defined in \p{CN}. 
The color factors come from contractions of the color tensors 
$$
\mathrm{Tr} ( t_{(a} \,t_{b)}) \;\; ,\;\; \mathrm{Tr} ( t_{(a} \,t_{b\vphantom{b)}} \,t_{c)}) \;\;,\;\;
\mathrm{Tr} ( t_{(a} \,t_{b\vphantom{b)}} \,t_{c\vphantom{c)}} \, t_{d)})
$$
appearing in the external vertices, with each other and with the antisymmetric structure constants $f_{abc}$  from the $\mathrm{T}$-blocks \p{Tblock}.

We also need to take into account the crossing symmetry of the correlator under the permutation of the external points $2 \rightleftarrows 3$ and of the insertion points $5 \rightleftarrows 6$. The latter corresponds to the exchange $\rho \rightleftarrows \sigma$. 
Thus we add to the  list of diagrams $\text{(E)}_{2 \rightleftarrows 3}$, 
and $\rho \rightleftarrows \sigma$ for (A),(B),$\text{(B)}_{2 \rightleftarrows 3}$,(C) and (D).
It is evident that the permutations do not alter the color and symmetry factors.

Then we sum up the contributions of all  diagrams and make use of the identities  (in the frame $\theta_5=\theta_6=0$)
$$
\rho_r^2 = \frac{(\theta^{+}_r)^2}{x_{r5}^2} \;\;,\;\; \sigma_r^2 = \frac{(\theta^{+}_r)^2}{x_{r6}^2} \;\; ,\;\;
\tau_{rs} = \frac{x_{rs}^2 x_{56}^2 (\theta^{+}_r)^2 (\theta^{+}_s)^2 }{x_{r5}^2 x_{r6}^2 x_{s5}^2 x_{s6}^2}  
$$
that enable us to explicitly identify the invariant $\Theta_{5,6}$ \p{Theta56}  in the expression for the two-loop correction to the correlator \p{Gg4}.
This yields
\begin{align*}
G_{g^4} = \frac{36 g^4 \mathcal{C}_N N^2}{(2 \pi)^{20}}\frac{1}{x_{12}^2 x_{13}^2 x_{24}^4 x_{34}^4}\frac{1}{x_{56}^6}
\Bigl[ x_{14}^2 x_{23}^2 x_{56}^2 + x_{12}^2 (x_{35}^2 x_{46}^2 + x_{36}^2 x_{45}^2) + x_{13}^2 (x_{25}^2 x_{46}^2 + x_{26}^2 x_{45}^2) + \\ +
x_{14}^2 (x_{25}^2 x_{36}^2 + x_{26}^2 x_{35}^2) + x_{23}^2 (x_{15}^2 x_{46}^2 + x_{16}^2 x_{45}^2) \Bigr]\,.
\end{align*}
Finally,  in order to apply the double insertion formula
$$
\mathcal{G}_{g^4}
= - \frac{1}{32} \int \mathrm{d}^4 x_5 \mathrm{d}^4 \theta_5 \int \mathrm{d}^4 x_6 \mathrm{d}^4 \theta_6\ 
 \mathcal{G}^{\mathrm{ins}}_{g^4}
$$
we switch the invariant $\Theta_{5,6}$ \p{Theta56} back to the frame $\theta^{+}_{1,2,3,4} = 0$. The $x$-space integrations give rise to the square of the one-loop ladder integral the and the two-loop ladder integrals from~\cite{Usyukina:1992jd}, as announced in \p{answer2}.

Let us note that we applied the tricks of identification of the harmonics just in order to perform the calculation in a concise way.
We could have kept all the harmonics different but  this results in a proliferation of Feynman diagrams. Some of them are constructed not only out of simple $\mathrm{T}$-blocks \p{Tblock} but also of the so-called double $\mathrm{T}$-blocks.  Moreover, each individual Feynman diagram depends 
on a number of harmonics $u^{-}$ that cancel out in the sum of all diagrams since the correlator has to respect H-analyticity 
at each perturbative order. The cancellation of the harmonics $u^{-}$ is a rather nontrivial property that partially fixes the relative numerical factors of the various 
diagrams and serves as a reliable check of the calculation. 


\acknowledgments

E.S. acknowledges a discussion with Pedro Vieira which stimulated this work. We are grateful to the authors of Ref.~\cite{Vieira:2013wya} for exchanging with us prior to publication. The work of D.C. has been supported by the ``Investissements d'avenir, Labex ENIGMASS''. He was partially supported by the RFBR grant 14-01-00341.

\end{document}